\documentclass[10pt]{article}
\usepackage{extsizes}
\usepackage[super,sort&compress,comma]{natbib} 
\usepackage[version=3]{mhchem}
\usepackage[left=2cm, right=2cm, top=3cm, bottom=2.0cm]{geometry}
\usepackage{balance}
\usepackage{times,mathptmx}
\usepackage{sectsty}
\usepackage{graphicx} 
\usepackage{lastpage}
\usepackage[format=plain,justification=justified,singlelinecheck=false,font={stretch=1.125,small,sf},labelfont=bf,labelsep=space]{caption}
\usepackage{float}
\usepackage{fancyhdr}
\usepackage{fnpos}
\usepackage[english]{babel}
\addto{\captionsenglish}{%
  
}
\usepackage{array}
\usepackage{droidsans}
\usepackage{charter}
\usepackage[T1]{fontenc}
\usepackage[usenames,dvipsnames]{xcolor}
\usepackage{setspace}
\usepackage[compact]{titlesec}
\usepackage{hyperref}

\usepackage{lipsum}
\usepackage{mathtools}
\usepackage{cuted}
\usepackage{subcaption}
\usepackage{widetext}

\definecolor{cream}{RGB}{222,217,201}

\begin{document}
\allowdisplaybreaks

\pagestyle{fancy}
\thispagestyle{plain}
\fancypagestyle{plain}{


\renewcommand{\headrulewidth}{0pt}
}

\makeFNbottom
\makeatletter
\renewcommand\LARGE{\@setfontsize\LARGE{15pt}{17}}
\renewcommand\Large{\@setfontsize\Large{12pt}{14}}
\renewcommand\large{\@setfontsize\large{10pt}{12}}
\renewcommand\footnotesize{\@setfontsize\footnotesize{7pt}{10}}
\makeatother

\renewcommand{\thefootnote}{\fnsymbol{footnote}}
\renewcommand\footnoterule{\vspace*{1pt}%
\color{cream}\hrule width 3.5in height 0.4pt \color{black}\vspace*{5pt}} 
\setcounter{secnumdepth}{5}

\makeatletter 
\renewcommand\@biblabel[1]{#1}            
\renewcommand\@makefntext[1]%
{\noindent\makebox[0pt][r]{\@thefnmark\,}#1}
\makeatother 
\renewcommand{\figurename}{\small{Fig.}~}
\sectionfont{\sffamily\Large}
\subsectionfont{\normalsize}
\subsubsectionfont{\bf}
\setstretch{1.125} 
\setlength{\skip\footins}{0.8cm}
\setlength{\footnotesep}{0.25cm}
\setlength{\jot}{10pt}
\titlespacing*{\section}{0pt}{4pt}{4pt}
\titlespacing*{\subsection}{0pt}{15pt}{1pt}


\makeatletter 
\newlength{\figrulesep} 
\setlength{\figrulesep}{0.5\textfloatsep}

\makeatother


\title{Entanglement via rotational blockade of MgF molecules in a magic potential}
\author{Eunmi Chae}
\date{Department of Physics, Korea University, Seoul, Republic of Korea \\ echae@korea.ac.kr}

\maketitle
\begin{center}
\begin{tabular}{p{13.5cm}}



\normalsize{Diatomic polar molecules are one of the most promising platforms of quantum computing due to their rich internal states and large electric dipole moments. Here, we propose entangling rotational states of adjacent polar molecules via a strong electric dipole-dipole interaction. The splitting of 1.27 kHz between two entangled states is predicted for MgF molecules in an optical tweezer array. The resolution of the entangled states can be achieved in a magic potential for the molecules where the rotational states have the same trap frequencies. The magic potential can be formed by tuning the angle between the molecules' quantization axis and the linear polarization of trapping light, so-called magic angle. We calculate the magic angle for MgF molecules in a reasonable experimental condition and obtain that the trap frequencies of the two involved states can be matched within a few 10s of Hz. Establishing entanglement between molecules, our results provide a first step towards quantum computing using diatomic polar molecules. } \\

\end{tabular}
\end{center}


\renewcommand*\rmdefault{bch}\normalfont\upshape
\rmfamily
\section*{}






\section{Introduction}
Diatomic polar molecules are a promising platform for quantum simulation and computing due to their rich internal structures and the strong electric dipole-dipole interaction. 
With their unique merits, molecules would enable brand-new experiments in quantum simulations and computations\cite{DeMille2002a, Carr2009, Baranov2012, Yan2013b, Wall2015a, Ni2018, Blackmore2019, Sawant2020, Hughes2020}, quantum chemistry\cite{Balakrishnan2016, Hu2019, Cheuk2020a, Liu2020a, Hu2020}, and precision measurements\cite{Hudson2011, Andreev2018}. 
Plentiful internal states of molecules can be employed for robust qubits with long coherence times, another knob for many body Hamiltonian, and various parity states for precision measurement. 
Various energy scales of molecular internal structures also open possibilities to couple the molecular qubits to other quantum platforms such as microwave technology for quantum communications\cite{Menzel2012, Brecht2016}. 
Moreover, polar molecules' strong electric dipole-dipole interaction enables simulations of strongly correlated systems with anisotropic long-range interactions as well as new methods for qubit gate operation. 
Molecules are also basic building blocks of chemical reactions. 
Controlling quantum states of molecules would allow quantum manipulations of chemical reactions.

To maximize their potential, preparation and coherent control of ultracold molecules are prerequisite.
With incessant efforts over the last decade, laser cooling and magneto-optical trapping (MOT) of diatomic molecules have been successfully realized to open a new era of ultracold molecules \cite{Barry2014, Anderegg2017, Truppe2017, Collopy2018}.
Soon after, sub-doppler cooling and trapping molecules in an optical tweezer array have been achieved so that quantum-level manipulations of individual molecules become within reach \cite{Truppe2017, Anderegg2018, Anderegg2019}.

Here, we propose quantum entanglement between polar molecules in an optical tweezer array by utilizing their rotational degrees of freedom and large electric dipole moments.
The two lowest rotational states can be treated as molecular qubits. 
The long lifetimes of the rotational states enable long coherence time which is crucial to guarantee robust manipulations of qubits by microwaves.  
When two molecules are in rotational states that have opposite parities with respect to each other, a strong electric dipole-dipole interaction between them can generate entanglement between the two molecular qubits. 

Magnesium monofluoride (MgF) molecules in an optical tweezer array are employed as our experimental platform. 
MgF is expected to have the best Franck-Condon factors among alkaline-earth monofluoride molecules\cite{Pelegrini2005, Xu2016}.
In addition, its strong transition at a short wavelength and its light mass make MgF the most efficient species that can be laser cooled. 
MgF has a rotational transition of 31 GHz and a large electric dipole moment of 3 Debye which can be utilized for qubit operations as well as simulating strongly correlated systems with long-range interactions.
Also, the molecule's three naturally occurring isotopes of both Bosons and Fermions position itself as an attractive platform for quantum simulation with different symmetries.

In this work, we present entanglement scheme of the rotational states of the two molecules by the electric dipole-dipole interaction between two MgF molecules in different rotational states. 
Also, conditions of a magic optical trap necessary for the entanglement is provided assuming reasonable experimental parameters for an optical tweezer array. 

\section{Molecular entanglement via rotational blockade}
\begin{figure}[h]
\centering
	\includegraphics[width=.6\textwidth]{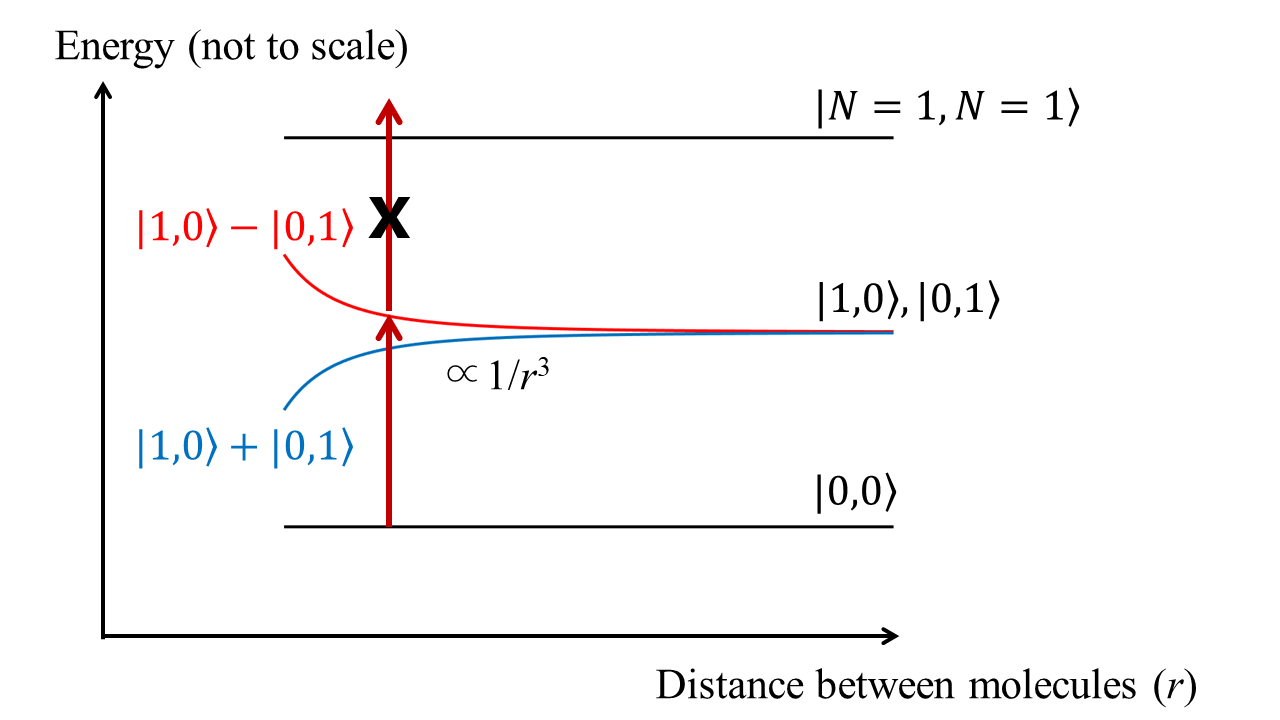}
  \caption{Rotational blockade. The electric dipole-dipole interaction causes $|1,0 \rangle$ and $|0,1 \rangle$ states to couple and become entangled states of $|1,0 \rangle + |0,1 \rangle$ and $|1,0 \rangle - |0,1 \rangle$. Here, the sign of the electric dipole-dipole interaction is assumed to be negative. When the electric dipole-dipole interaction is larger than the linewidth of transitions, a microwave that is resonant to the entangled states cannot excite the molecules to the $|1,1 \rangle$ state.}
  \label{fg:DDI}
\end{figure}

The electric dipole-dipole interaction takes place when the two involved states have the opposite parities.
We employ two rotational states, the rotational ground state ($N=0$, $N$: molecule's rotational quantum number) with the even parity and the first excited rotational state ($N=1$) with the odd parity, to form a molecular qubit.
When two adjacent molecules are both in the $N=0$ states or in the $N=1$ states (denoted as $|0,0 \rangle$ and $|1,1 \rangle$ respectively), there is no dipole-dipole interaction between them.
The electric dipole-dipole interaction acts on the two molecules when only one of the molecules is excited to the $N=1$ state ($|0,1 \rangle $ or $|1,0 \rangle $ states).
As a result, $|0,1 \rangle$ and $|1,0 \rangle$ states, who are degenerate without the electric dipole-dipole interaction, split to $|0,1 \rangle + |1,0 \rangle$ and $|0,1 \rangle-|1,0 \rangle$ states where the two molecules are entangled (Fig \ref{fg:DDI}).
The energy of the splitting depends on the electric dipole moments of the molecules and the distance between them. 
The two molecules become entangled by a microwave whose frequency is tuned to be resonant to one of the coupled states.
A two photon transition from $|0,0 \rangle$ to $|1,1 \rangle$ does not occur when the energy splitting due to the electric dipole-dipole interaction is larger than the linewidth of the transition as shown in Fig. \ref{fg:DDI}. 
We call this effect as "rotational blockade".
Rotational blockade can entangle multiple molecules within the length where the electric dipole-dipole interaction is in effect, resulting interesting phase transitions depending on the distance of molecules as in Rydberg atoms.
Moreover, due to the anisotropy of the electric dipole-dipole interaction, exotic phases may arise depending on the angle between the positions of molecules and the molecular quantization axis \cite{Wall2015a}. 

The electronic ground state of $^{24}$Mg$^{19}$F is a type B molecule whose wavefunction is expressed by quantum numbers of rotation $N$, total electronic angular momentum $J=N+S$, and total angular momentum $F=J+I$.
The electron spin $S$ and the nuclear spin $I$ of MgF are both 1/2. 
Among the hyperfine states, we employ $|N=0,J=1/2,F=0,M_F=0 \rangle$ state and $|N=1, J=3/2, F=1, M_F=0 \rangle$ state as the two qubit states in our setup. 
They are denoted as $|\downarrow \rangle$ and $|\uparrow \rangle$ respectively in the following discussion. 

The Hamiltonian of the electric dipole-dipole interaction is
\begin{equation}
 	H_{\textrm{dd}} = -\frac{1}{4 \pi \epsilon_0 r^3} \left[ \frac{3 (\hat{\mathbf{d}}^1 \cdot \mathbf{r}) (\hat{\mathbf{d}}^2 \cdot \mathbf{r})}{r^2} -\hat{\mathbf{d}}^1 \cdot \hat{\mathbf{d}}^2 \right]
\end{equation}
, where $\hat{\mathbf{d}}^i$ indicates the electric dipole operator of the $i$th molecule and $\mathbf{r}$ is the separation between the two molecules.
Among components of this Hamiltonian, only those that do not change the total angular momentum projection quantum number $M_F$ have non-zero values when the v$M_F$ values of the two involved states are the same value as in our case.
As the angle between $\mathbf{r}$ and the molecules' quantization axis is defined as $\theta$, the relevant Hamiltonian is 
\begin{equation}
	H_{\textrm{dd}, \, q=0} = \frac{1-3 \cos^2 \theta}{4 \pi \epsilon_0 r^3} \left[ {\hat{d}_0}^1 {\hat{d}_0}^2 + \frac{1}{2} \left( {\hat{d}_1}^1 {\hat{d}_{-1}}^2 + {\hat{d}_{-1}}^1 {\hat{d}_1}^2 \right) \right]
\end{equation}
, where ${\hat{d}_{\pm 1}}^i = \mp ({\hat{d}_x}^i \pm i {\hat{d}_y}^i)/\sqrt{2}$ and ${\hat{d}_0}^i = {\hat{d}_z}^i$ are spherical components of the dipole operator on the $i$th molecule, and $q$ is the angular momentum projection of the electric dipole-dipole interaction. 

The matrix elements of the dipole operator can be calculated as below \cite{Brown2003}.

\begin{widetext}
\begin{eqnarray}
	\langle N', J', F', M_F' | \hat{d}_p | N, J, F, M_F \rangle &=& (-1)^{F'-M_F'} 
	\begin{pmatrix}
		F' & 1 & F \\
		-M_F' & p & M_F
	\end{pmatrix} 
	\langle N', J', F' || \hat{\mathbf{d}} || N, J, F \rangle  \label{eq:dipole1} \\
	\langle N', J', F' || \hat{\mathbf{d}} || N, J, F \rangle &=& (-1)^{F'+J+1+I} \sqrt{(2F'+1)(2F+1)} 
	\begin{Bmatrix}
		F' & J' & I' \\
		J & F & 1
	\end{Bmatrix}
	\langle N', J' || \hat{\mathbf{d}} || N, J \rangle \label{eq:dipole2} \\
	\langle N', J' || \hat{\mathbf{d}} || N, J \rangle &=& (-1)^{J' + N + 1 + S} \sqrt{(2J'+1)(2J+1)}
	\begin{Bmatrix}
		J' & N' & S' \\
		N & J & 1
	\end{Bmatrix} 
	\langle N' || \hat{\mathbf{d}} ||N \rangle \label{eq:dipole3} \\
	\langle N' || \hat{\mathbf{d}} ||N \rangle &=& d (-1)^{N' - \Lambda'} \sqrt{(2N'+1)(2N+1)}
	\begin{pmatrix}
		N' & 1 & N \\
		-\Lambda' & 0 & \Lambda
	\end{pmatrix} \label{eq:dipole4}
\end{eqnarray}
\end{widetext}
$\langle N', J', F' || \hat{\mathbf{d}} || N, J, F \rangle$, $\langle N', J' || \hat{\mathbf{d}} || N, J \rangle$, $\langle N' || \hat{\mathbf{d}} || N \rangle$ are reduced matrix elements in each corresponding subspace. 
$\Lambda$ indicates the projection of rotation ($N$) to the molecule's internuclear axis and $\Lambda = 0$ for the electronic ground state of MgF.
In the subspace of our interest, the matrix elements of $\hat{d}_{\pm 1}$ are zero since $M_F = 0$ for both of the two involved states $|\downarrow \rangle $ and $|\uparrow \rangle$. 
The matrix elements of $\hat{d}_0$ in our subspace are calculated by substituting the quantum numbers of the involved states to the equations (\ref{eq:dipole1}) $\sim$ (\ref{eq:dipole4}).
\begin{eqnarray}
	\langle \downarrow |\hat{d}_0|\downarrow \rangle &= \langle \uparrow|\hat{d}_0|\uparrow \rangle &= 0  \\
	\langle \downarrow |\hat{d}_0|\uparrow \rangle &= \langle \uparrow |\hat{d}_0|\downarrow \rangle &= \frac{\sqrt{2}}{3}d
\end{eqnarray}
Here, $d = 3.077$ Debye is the electric dipole moment of the electronic ground state of the MgF molecule \cite{Langhoff1986}.
Therefore, the total Hamiltonian of our molecular qubit system is expressed as
\begin{equation}
	H = 
	\begin{pmatrix}
		2E_0 & 0 & 0 & 0 \\
		0 & E_0 & \frac{1-3\cos^2 \theta}{4 \pi \epsilon_0 r^3} \frac{2}{9} d^2 & 0 \\
		0 & \frac{1-3\cos^2 \theta}{4 \pi \epsilon_0 r^3} \frac{2}{9} d^2 & E_0 & 0 \\
		0 & 0 & 0 & 0
	\end{pmatrix}
\end{equation}
in the basis of $|\uparrow,\uparrow \rangle, |\uparrow,\downarrow \rangle, |\downarrow,\uparrow \rangle$, and $|\downarrow,\downarrow \rangle$.
We set the energy of the $| \downarrow \rangle$ state as zero. 
$E_0=30.99$ GHz is the energy difference between $|\downarrow \rangle$ and $|\uparrow \rangle$ states of MgF \cite{Anderson1994a}.
By diagonalizing this Hamiltonian, one can find that the two states $|\uparrow,\downarrow \rangle$ and $|\downarrow,\uparrow \rangle$ are coupled by the electric dipole-dipole interaction to form $\left( |\uparrow,\downarrow \rangle + |\downarrow,\uparrow  \rangle \right)/\sqrt{2} $ and $\left( |\uparrow,\downarrow \rangle - |\downarrow,\uparrow  \rangle \right)/\sqrt{2} $ states where the internal states of the two molecules are entangled. 
The splitting of the two coupled states is twice of the strength of the dipole-dipole interaction, which takes the maximum value when the molecules' quantization axis is parallel to the displacement between the two molecules ($\theta = 0$). 
The splitting is estimated to be 1.27 kHz in experimental conditions of $r = 1$ $\mu$m and $\theta = 0$.
The entangled state can be generated using a microwave that is resonant to one of the coupled states. 
When the electric dipole-dipole interaction is large enough compared to the linewidth of the transition, the probability for the two molecules to be simultaneously excited to the $|\uparrow,\uparrow \rangle$ state is very low since the two photon transition is out-of-resonance (Fig. \ref{fg:DDI}).
We call this effect "rotational blockade".

\section{Magic angle of optical tweezers for MgF}
The total linewidth of the rotational transitions should be narrower than the splitting due to the electric dipole-dipole interaction in order to resolve the two entangled states. 
Since the natural linewidth of the rotational transitions is extremely narrow, the total linewidth is determined by the experimental conditions.
For molecules in optical tweezers, different lightshifts of the two involved rotational states are the dominant source of the linewidth broadening.
A light with 532 nm wavelength, 1 mW power, and 1 $\mu$m $1/e^2$ diameter generates a trap for a MgF molecule in the $|\downarrow \rangle$ state with the trap frequency of 190 kHz in the radial direction and 41 kHz in the propagation direction.
At 30 $\mu$K which is the lowest temperature achieved so far with CaF molecules in optical tweezers \cite{Anderegg2019a}, about 380 motional states can be populated with more than 10\% probability. 
This motional degree of freedom can result in linewidth broadening of about 100 kHz which is much larger than the strength of the electric dipole-dipole interaction between the molecules. 
Therefore, the polarizabilities of the $|\downarrow \rangle$ and $|\uparrow \rangle$ states of MgF should be adjusted in order to match the trap frequencies of optical tweezers for them. 

\begin{figure}[h]
\centering
  \includegraphics[width=.5\textwidth]{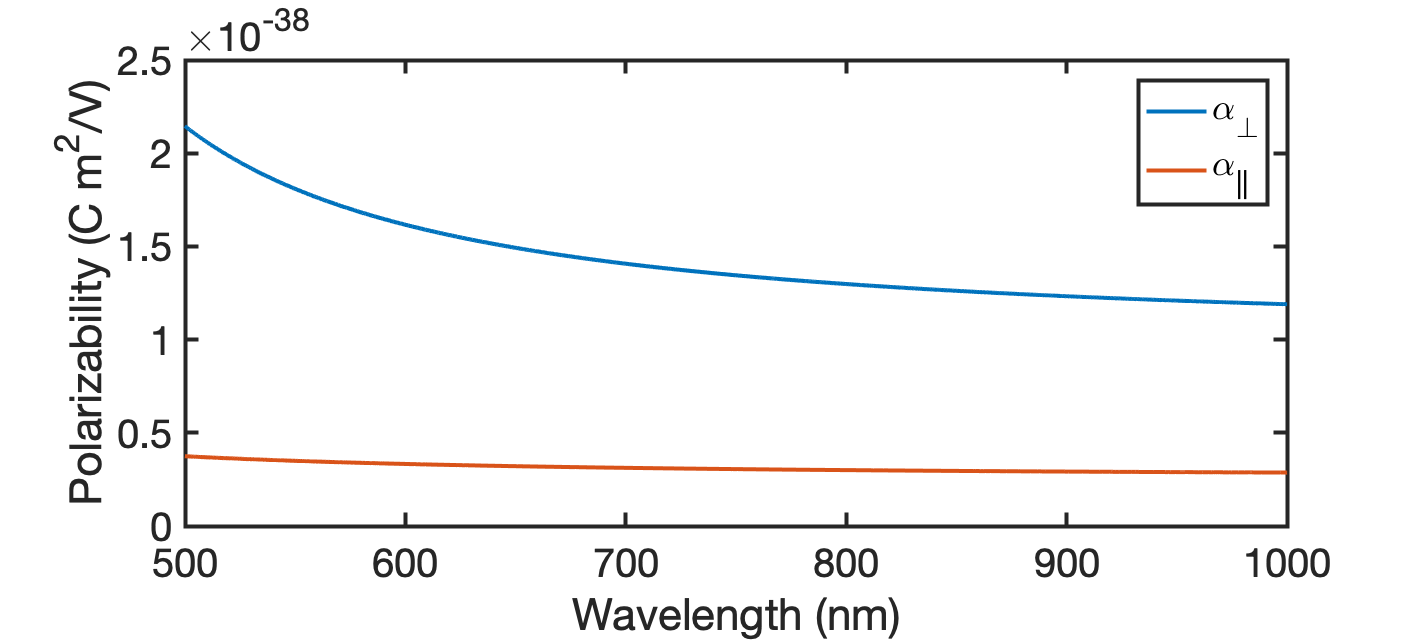}
  \caption{$\alpha_{\parallel}$ and $\alpha_{\perp}$ of the electronic ground state of MgF molecules. }
  \label{fg:polarizability}
\end{figure}

Diatomic polar molecules have two types of polarizabilities – one parallel to the molecule's internuclear axis ($\alpha_{\parallel}$) and the other perpendicular to the axis ($\alpha_{\perp}$).
Transitions from a $\Sigma$ state to another $\Sigma$ state such as the $X$-$B$ transition of MgF contribute to $\alpha_{\parallel}$.
$\alpha_{\perp}$ is determined by transitions from a $\Sigma$ state to a $\Pi$ state, such as the $X$-$A$ transition of MgF.
The two polarizabilities at visible and near infra-red wavelengths can be calculated from the transition frequencies and strengths of the $X$-$A$ and $X$-$B$  transitions as below \cite{Bonin1997}.
\begin{eqnarray}
	\alpha_{\parallel} &=& \frac{2\omega_{XB}}{\hbar (\omega_{XB}^2 - \omega^2)} | \langle X | \hat{d}_0^{\mathrm{mol}} | B \rangle |^2 \\
	\alpha_{\perp} &=& \frac{2\omega_{XA}}{\hbar (\omega_{XA}^2 - \omega^2)} | \langle X | \hat{d}_{-1}^{\mathrm{mol}} | A \rangle |^2 
\end{eqnarray}
$\omega_{XA}$ and $\omega_{XB}$ are the resonant transition frequencies of $X-A$ and $X-B$ transitions respectively. 
$\hat{\mathbf{d}}^{\mathrm{mol}}$ is the electric dipole operator in the molecular frame. 
The contributions from higher states such as $X$-$C$ and $X$-$D$ transitions are negligible at this wavelength since their transition frequencies are deep in ultra-violet\cite{Novikov1971}. 
The calculated polarizabilites using the wavelengths and linewidths of the $X-A$ and $X-B$ transitions of MgF are shown in Fig. \ref{fg:polarizability} \cite{Pelegrini2005, Yang2017}.
In our setup, we are planning to use a wavelength of 532 nm for the optical tweezers. 

\begin{figure}[h]
\centering
  \includegraphics[width=.7\textwidth]{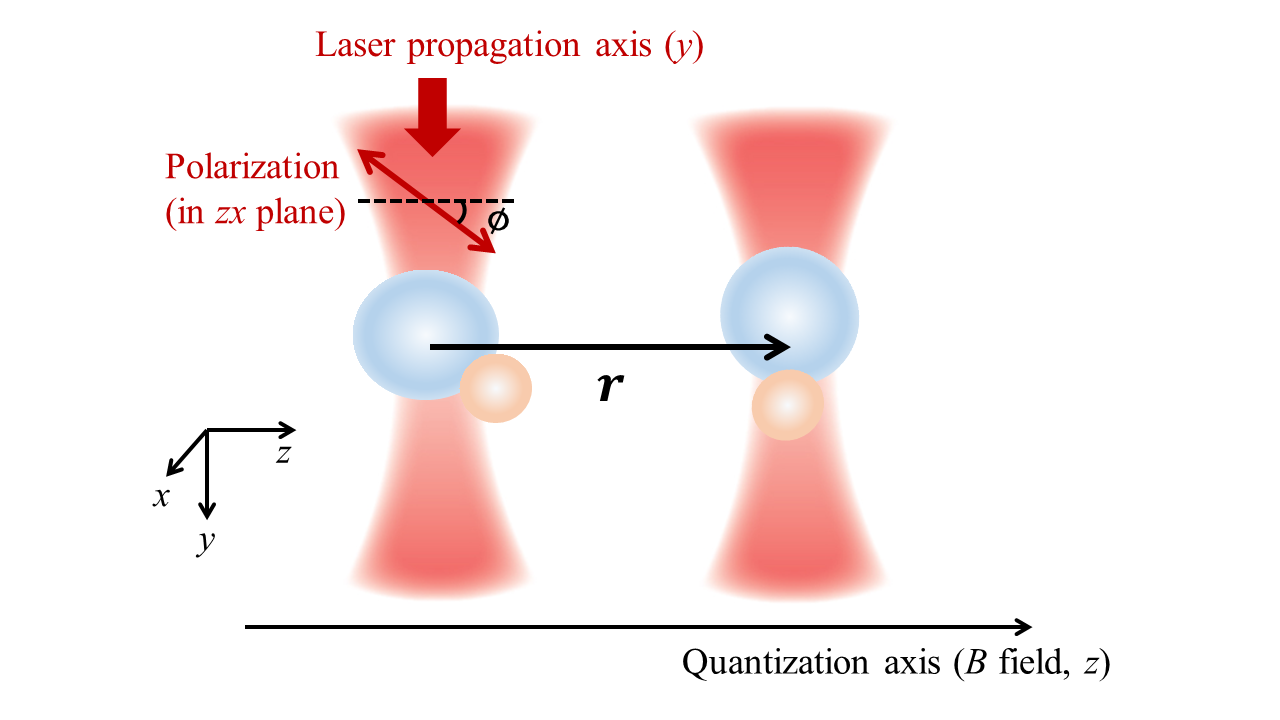}
  \caption{Experimental setup for molecular qubits. The molecules' quantization axis is set to be aligned to the tweezer array using a magnetic field to maximize the electric dipole-dipole interaction. The angle between the quantization axis and the polarization of trapping light is denoted as $\phi$. }
  \label{fg:layout}
\end{figure}

The effective polarizability of the molecule is a weighted sum of the two polarizabilities depending on the internal state of the molecule and the direction of the linear polarization of the optical trap\cite{Bonin1997}. 
In the lab frame, the rotational ground state ($N=0$) has a spherical wavefunction.
Therefore, its polarizability does not depend on the direction of the polarization of the trapping light.
On the other hand, the molecular wavefunction of the $N=1$ state has an elongated p-wave-like shape and the elongated direction is determined by the projection quantum number of rotation ($M_N$) in the lab frame.
Therefore, the angle between the quantization axis and the linear polarization of the trapping light ($\phi$ in Fig. \ref{fg:layout}) determines the effective polarizability \cite{Neyenhuis2012}. 
In the experiment, the quantization axis is set by a magnetic field parallel to the molecular separation in order to maximize the electric dipole-dipole interaction.
By changing the direction of the linear polarization of the trapping laser with respect to the magnetic field, we can adjust the polarizability of the $N=1$ state so that it matches to the one of the $N=0$ state. 

We need to evaluate an effective polarizabilities of our qubit states in the lab frame in order to calculate the optical trap depth and frequencies.
The interaction between the light and the molecule affects the rotation quantum number $N$, $M_N$, but it does not change the electron spin $S$, $M_S$ and the nuclear spin $I$, $M_I$ of the molecule.
Therefore, our qubit states which are described by the total angular momentum $|N, (S), J, (I), F, M_F \rangle$ should be decoupled to $|N, M_N \rangle |S, M_S \rangle |I, M_I \rangle$ representations in order to evaluate their polarizabilities.

It is convenient to separate a scalar part of the polarizability $\alpha_{\mathrm{sc}}$ and a tensor part $\alpha_{\mathrm{t}}$ which are expressed using the molecular-frame polarizabilities as below \cite{Bonin1997}.
\begin{eqnarray}
	\alpha_{\mathrm{sc}} &=& \frac{1}{3} \left( \alpha_{\parallel} + 2\alpha_{\perp} \right) \\
	\alpha_{\mathrm{t}} &=& \frac{2}{3} \left( \alpha_{\parallel} - \alpha_{\perp} \right) 
\end{eqnarray}
The polarizability of the $|\downarrow \rangle$ state which is one of the $N=0$ states has only a scalar component $\alpha_{\mathrm{sc}}$.
On the other hand, we need to consider the contribution of the tensor part $\alpha_{\mathrm{t}}$ for the $|\uparrow \rangle$ state.

When the linear polarization of trapping light points to the quantization axis $z$ ($\phi=0$), the Hamiltonian of the optical trap for the $N=1$ state can be written as
\begin{equation}
	H_{\mathrm{trap},\, N_{zxy}} = -\frac{1}{c \epsilon_0} 
	\begin{pmatrix}
		b & 0 & 0 \\
		0 & a & 0 \\
		0 & 0 & a
	\end{pmatrix}
	I
\end{equation}
 in $|N=1,M_N=z,x,y \rangle$ basis.
Here, $I$ is the intensity of trapping light, and $a = \alpha_{\mathrm{sc}} - 1/5 \alpha_{\mathrm{t}}$ and $b = \alpha_{\mathrm{sc}} + 2/5 \alpha_{\mathrm{t}}$ are the effective polarizabilities of the $|N=1, M_N=x,y \rangle$ states and the $|N=1, M_N = z \rangle$ state respectively \cite{Neyenhuis2012}.
One can derive the Hamiltonian of the optical trap for $|N=1, J=3/2, F=1, M_F=z, x, y \rangle$ states by expressing the states in decoupled representations.
The resulting Hamiltonian is
 \begin{equation}
	H_{\mathrm{trap},\, F_{zxy}} = -\frac{1}{c \epsilon_0} 
	\begin{pmatrix}
		\frac{1}{3} (a + 2b) & 0 & 0 \\
		0 & \frac{1}{6}(5a + b) & 0 \\
		0 & 0 & \frac{1}{6}(5a + b)
	\end{pmatrix}
	I
\end{equation}
. 

This Hamiltonian needs to be rotated when the polarization of trapping light has an angle $\phi \neq 0$ to the quantization axis as shown in Fig. \ref{fg:layout}.
The resulting Hamiltonian is 
\begin{equation}
	H_{\mathrm{trap}, \, F_{zxy},\, \phi} = R^{\mathrm{T}} H_{\mathrm{trap},\, F_{zxy}} R
\end{equation}
where,
\begin{equation}
	R = 
	\begin{pmatrix}
		\cos \phi & -\sin \phi & 0 \\
		\sin \phi & \cos \phi & 0 \\
		0 & 0 & 1
	\end{pmatrix}
\end{equation}
is a rotation matrix around $y$ axis in the $z, x, y$ basis.

\begin{figure}[h]
\centering
  \includegraphics[width=.5\textwidth]{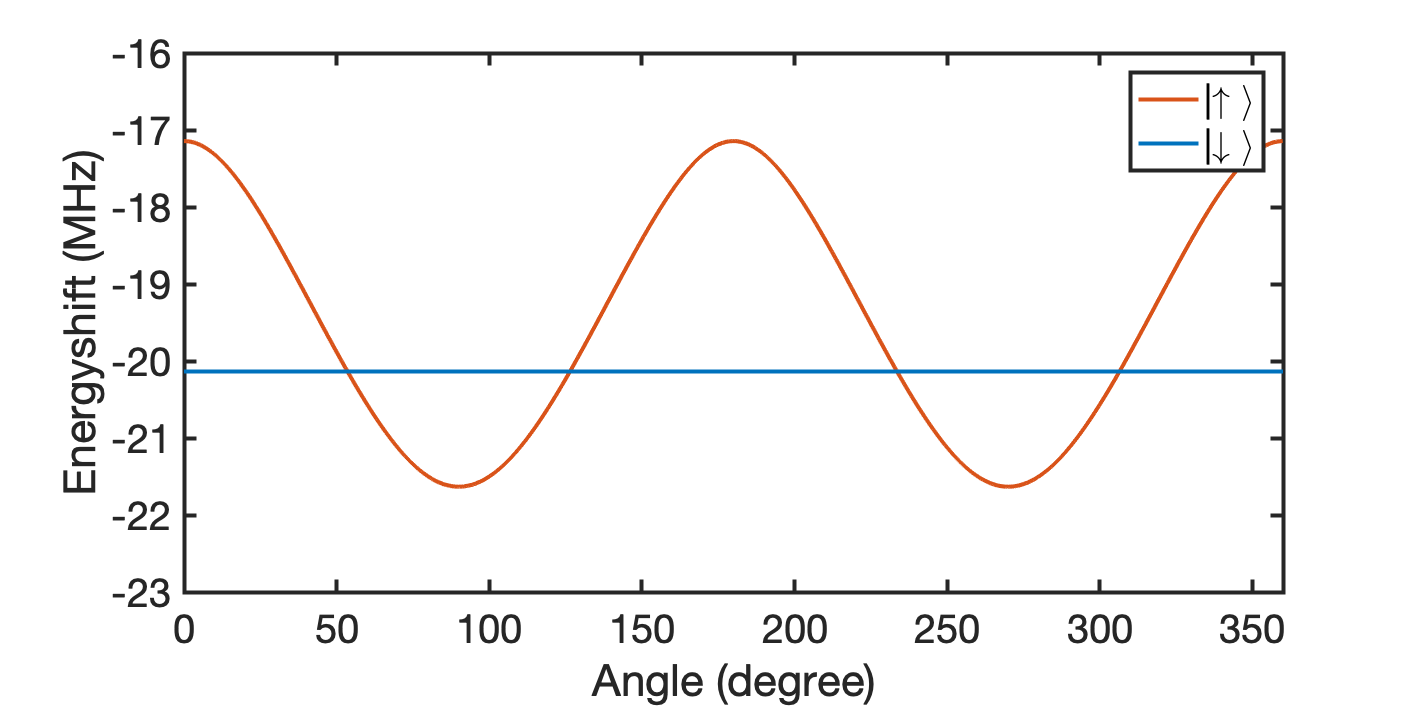}
  \caption{Optical trap depth of the $|\downarrow \rangle$ and $|\uparrow \rangle$ states at the magnetic field of 10 Gauss. The power of the 532 nm trapping light is set to be 1 mW and the $1/e^2$ diameter is 1 $\mu$m. The depth of the $|\uparrow \rangle$ state changes as the polarization of the light field is rotated while the depth of the $|\downarrow \rangle$ state remains constant.}
  \label{fg:Angledependence}
\end{figure}

It is convenient to express the total Hamiltonian in $|F=1, M_F = 0, -1, 1 \rangle$ basis ($H_{\mathrm{trap},\, F,\, \phi}$) in order to accommodate the Zeeman energyshift due to the quantizing magnetic field.
The total Hamiltonian is the sum of the optical trapping Hamiltonian and the energy of each magnetic substate at the given magnetic field.
\begin{equation}
	H_{\mathrm{total}} = H_{\mathrm{trap}, \, F, \, \phi} + 
	\begin{pmatrix}
		E_0 & 0 & 0 \\
		0 & E_{-1} & 0 \\
		0 & 0 & E_1
	\end{pmatrix}
\end{equation}
The lightshift of the qubit state $|\uparrow \rangle = |N=1, J=3/2, F=1, M_F = 0 \rangle$ can be obtained by diagonalizing this Hamiltonian at each angle $\phi$. 
The orange line in Fig. \ref{fg:Angledependence} shows the result for the trapping laser of 532 nm wavelength, 1 mW power, and 1 $\mu$m $1/e^2$ beam diameter. 
The quantizing magnetic field is set to be 10 Gauss along $z$ axis. 
The trap depth of the $|\uparrow \rangle$ state changes more than 20\% as the polarization of the light field is rotated with respect to the quantization axis. 
By contrast, the blue line in Fig. \ref{fg:Angledependence} indicates the constant trap depth of the $|\downarrow \rangle$ state.
By adjusting the polarization angle of the trapping laser, the trap frequency of the $|\uparrow \rangle$ state can be matched to the one of the $|\downarrow \rangle$ state.
This angle is called as magic angle \cite{Neyenhuis2012}.
The magic angle for our experimental parameters is estimated to be 51.4 degrees. 

\begin{figure}[h]
	\centering
	\includegraphics[width=0.8\textwidth]{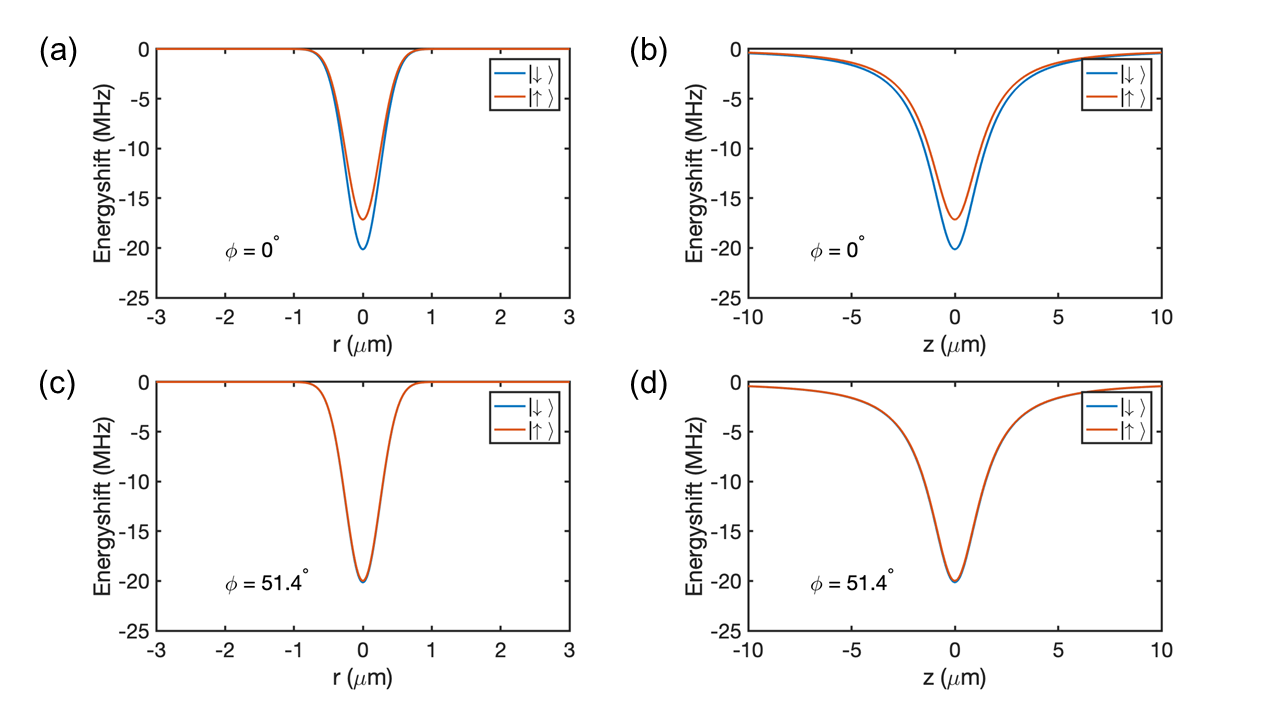}
	\caption{Optical traps of the $|\downarrow \rangle $ and $|\uparrow \rangle$ states for two different angles between the quantization axis and the polarization of the optical trap. The trapping laser at 532 nm is assumed to have the power of 1 mW with the $1/e^2$ diameter of 1 $\mu$m. The quantization magnetic field is 10 Gauss. (a),(b) $\phi = 0^{\circ}$. (c),(d) $\phi = 51.4^{\circ}$.}
	\label{fg:opticaltraps}
\end{figure}

Fig. \ref{fg:opticaltraps} depicts the optical trap depth for the $|\downarrow \rangle$ and $|\uparrow \rangle$ states in the above experimental conditions with two different values of angle $\phi$. 
When the polarization is aligned to the quantization axis ($\phi = 0$), the two states feel different trap depths and the difference between the trap frequencies are 15 kHz (radial direction) and 3.2 kHz (propagation direction) (Fig. \ref{fg:opticaltraps} (a),(b)).
By tuning the polarization of the trapping light ($\phi = 51.4^{\circ}$, magic angle), the differences between the trap frequencies of the $|\downarrow \rangle$ and $|\uparrow \rangle$ states are reduced dramatically down to 17 Hz (radial direction) and 29 Hz (propagation direction) (Fig. \ref{fg:opticaltraps} (c),(d)).
Molecular entanglement by rotational blockade becomes within reach at this condition since the linewidth broadening due to the motional degrees of freedom becomes smaller than the electric dipole-dipole interaction between the $|\downarrow \rangle$ and $|\uparrow \rangle$ states.

\section{Conclusions}
In this work, we propose an entanglement scheme for molecular qubits in an optical tweezer array via rotational blockade. 
The rotational states of polar molecules with long coherence time and high controllability are employed as our molecular qubit states. 
The strong electric dipole-dipole interaction between two rotational states with opposite parities entangles the rotational states of adjacent molecules in optical tweezers. 
We calculate the energy splitting of the entangled states ($|\downarrow, \uparrow \rangle + |\uparrow, \downarrow \rangle$ and $|\downarrow, \uparrow \rangle - |\uparrow, \downarrow \rangle$) to be upto 1.27 kHz when two MgF molecules are separated by 1 $\mu$m. 
However, rotational blockade may be hindered by the difference in lightshifts of the $|\downarrow \rangle$ and the $|\uparrow \rangle$ states which is estimated to be about 100 kHz in a typical experimental condition. 
This problem can be overcome when the lightshifts of the two states are matched by tuning the angle between molecules' quantization axis and the linear polarization of the trapping light. 
Our calculation shows the difference in the trapping frequencies of the two qubit states of MgF can be reduced down to 17 Hz and 29 Hz along radial and propagation direction respectively in a reasonable experimental condition.  
This work guides an experimental realization of the molecular entanglement which is the crucial first step toward the quantum computation using polar molecules. 

\section*{Conflicts of interest}
There are no conflicts to declare.

\section*{Acknowledgements}
The author would like to thank Donghyun Cho and Kang-Kuen Ni for the valuable discussions and comments. This work was supported by the National Research Foundation of Korea (Grant No. 2020R1A4A101801511, 2020R1F1A107416211).


\balance


\bibliography{References-RotationalBlockade} 
\bibliographystyle{rsc} 

\end{document}